\documentclass[pra,twocolumn,epsfig,rotate,showpacs]{revtex4}
\usepackage{amsfonts} 
\usepackage{amsmath}
\usepackage{amssymb}
\usepackage{bm}
\usepackage[usenames, dvipsnames]{color} 
\usepackage{dcolumn}
\usepackage{epic} 
\usepackage{epsfig}
\usepackage{graphicx}
\usepackage[abs]{overpic}
\usepackage[lofdepth,lotdepth,caption=false]{subfig}

\usepackage[breaklinks,colorlinks = true,linkcolor = blue,urlcolor=blue,citecolor=blue]{hyperref}
\usepackage{soul}
\usepackage{times}
\usepackage{ulem}
\usepackage{wrapfig} 
\usepackage{xy} 

\newcommand{\beq}{\begin{eqnarray}}
\newcommand{\eeq}{\end{eqnarray}}

\newcommand{\dg}{\dagger}

\newcommand{\la}{\langle}
\newcommand{\ra}{\rangle}

\begin{document}
\title{A Simple Hubbard Model for the Excited States of Dibenzoterrylene}
\author{Z. S. Sadeq}
\email{sadeqz@physics.utoronto.ca}
\author{J.E. Sipe}
\affiliation{Department of Physics, University of Toronto, 60 St. George Street, Toronto, Ontario, Canada, M5S 1A7}
\date{\today}
\begin{abstract}
We use a simple Hubbard model to characterize the electronic excited states of the dibenzoterrylene (DBT) molecule; we compute the excited state transition energies and oscillator strengths from the ground state to several singlet excited states. We consider the lowest singlet and triplet states of the molecule,  examine their wavefunctions, and compute the density correlation functions that describe these states. We find that the DBT ground state is mostly a closed shell singlet with very slight radical character. We predict a relatively small singlet-triplet splitting of 0.75 eV, which is less than the mid-sized -acenes but larger than literature predictions for this state; this is because the Hubbard interaction makes a very small correction to the singlet and triplet states. 


\end{abstract}
\email{sadeqz@physics.utoronto.ca}
\maketitle

\section{Introduction}

Single photon sources are an important resource for quantum information processing \cite{frZPL,singp,zv1,zv2,nico}. Candidate devices are based on semiconductor quantum dots, color centers in diamonds, and trapped atoms or ions in the gas phase \cite{frZPL}. Organic dye molecules, at cryogenic temperatures, also can act as a source of single photons, with the optical coherence lifetime of the relevant transitions longer by an order of magnitude than those of semiconductor quantum dots \cite{frZPL, DBTref1, DBTref2,DBTref3}. Synthesis of these molecules is relatively straightforward and they are simple to deposit on optical chips and waveguides, thereby opening up the possibility of using existing integrated chip strategies to carry out a variety of nonlinear optical processes \cite{kylem}. 

\begin{figure}[h]
\begin{center}
\includegraphics[scale=0.45]{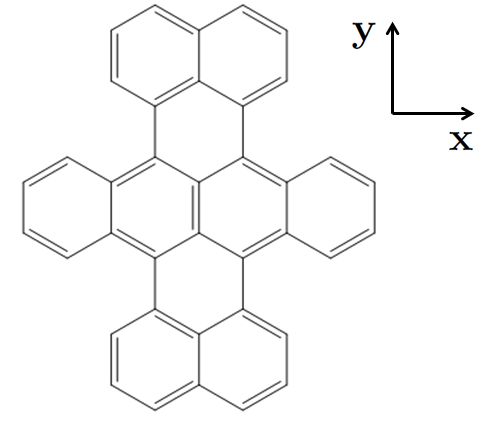}
\caption{A cartoon representation of the DBT molecule with the hydrogen atoms not shown along with the two molecular axes.}
\label{dbtmolfig}
\end{center}
\end{figure}

Dibenzoterrylene (DBT) is one of these dye molecules. It is composed of benzene rings; a cartoon representation of DBT is shown in Fig \ref{dbtmolfig}. Typically, these molecules are deposited in an anthracene (Ac) matrix, primarily to guard against oxidation and photobleaching, processes that limit the photostability of the molecule. DBT has a purely electronic, zero phonon line (ZPL) around 785 nm. At low temperatures, the phonon induced dephasing of the transition dipole of the ground to $S_{1}$ state vanishes; the spectral line width of this transition is then limited only by the electronic dephasing time, and DBT can act as a two level system, similar to a trapped atom \cite{frZPL}.

The development of such molecules as single photon sources requires an understanding of their electronic states \cite{depr10}. There are multiple strategies for calculating these, such as Density Function Theory (DFT) \cite{depr10,dftcalc2} and applications of Density Matrix Renormalization Group (DMRG) \cite{PPP,hach}. These techniques can be quite accurate, but are often computationally intensive. DMRG, in particular, is a technique that has yielded significant insight into the nature of the ground state of the -acene series; the  mid sized -acenes, such as tetracene, pentacene, and hexacene, are a set of molecules closely related to DBT in their structure. Hachmann \textit{et al.} have found that the ground state of large -acenes tend to be polyradical in nature \cite{hach}. 

An important drawback to using organic materials as single photon sources is intersystem crossing (ISC), a process where the excitation in the first singlet state, $S_{1}$ is funneled to the first triplet state, $T_{1}$ \cite{orgpc}. The rate at which ISC proceeds is proportional to the inverse of the energy gap between the singlet and triplet state, and therefore the energy of the triplet state is important in considering the use of DBT as a single photon source. There is some controversy as to the energy of the first triplet state of DBT; some have speculated this triplet state to be as low in energy as 0.23 eV  above the ground state\cite{depr10}. This energy is very different from that of the triplet states in the -acenes, which are typically on the order of 1 eV above the ground state. 

Earlier \cite{ZSpen} we presented a strategy which uses the Hubbard model to describe the electronic structure of the mid sized -acenes, with reasonable quantitative agreement with experiment; our approach can also take into account the polyradical nature of the ground state in these conjugated -acene systems.  As well as giving predictions for the transition energies of these molecules, our model allows us to compute oscillator strengths for the transitions, and the charge density correlation function for electronic states of interest.  

In this paper, we use this model to elucidate the electronic excited states of the DBT molecule; the attractive feature of our model is its ability to provide a simple physical picture of the electron behavior in these states. We show reasonable agreement with some DFT calculations, and make a prediction for the experimentally elusive triplet state that yields a higher value for the triplet energy than an earlier calculation \cite{depr10}. Our value is more in line with those of the mid sized -acenes, which are similar in size and structure to DBT. We also compute natural orbital occupation in DBT and compare it with the -acenes. 

As in our previous work \cite{ZSpen}, we characterize the excited states by computing the normalized electron density correlation function $g^{(2)}(s)$. We analyze the electron behavior in the first singlet and triplet excited states, and discuss the singlet-triplet gap in DBT versus the -acenes.  

This paper is written in five parts. In Section \ref{mod} we discuss the model used to describe the electronic states of DBT, in Section \ref{eos} we list the energies and oscillator strengths of the singlet and triplet excited states, in Section \ref{sos} we discuss the lowest singlet and triplet excited states, and in Section \ref{conc} we conclude. 
\section{\label{mod} Model}
We use a simple Hubbard model with a limited basis to describe the electronic states of the DBT molecule. The Hamiltonian consists of a tight binding contribution and a Hubbard contribution,
\begin{equation}
H = H_{TB} + H_{Hu},
\label{fulham}
\end{equation}
where the tight binding Hamiltonian is
\begin{equation}
H_{TB} = - \sum_{\langle i,j \rangle, \sigma} t_{ij} c^{\dagger}_{i\sigma} c_{j\sigma}.
\label{TBeq}
\end{equation}
Here $\sigma$ is a spin label and $i,j$ are site labels; the angular brackets indicates sum over nearest neighbors, and the operators $c_{i\sigma}$ obey fermionic anti-commutation relations, $\{ c_{i\sigma},c^{\dagger}_{j\sigma'}\} = \delta_{ij}\delta_{\sigma\sigma'}$. We take the hopping parameter, $t_{ij}$ to be $t_{D} = 2.66 \text{ eV}$ for the $\pi$-conjugated bonds, and to be $t_{S} = 2.22 \text{ eV}$ for the single bonds, in accordance with literature values \cite{barf,cpam,cpam2}.

The use of the tight binding Hamiltonian (\ref{TBeq}) leads to a degeneracy between singlet and triplet states; this is broken by the electron-electron repulsion. The easiest way to take this into account is by a simple Hubbard contribution of the form
\begin{equation}
H_{Hu} = U \sum_{i} n_{\uparrow}(i) n_{\downarrow}(i),
\label{HubHam}
\end{equation}
where $n_{\sigma}(i)$ is the number operator, defined as $n_{\sigma} (i) \equiv  c^{\dagger}_{i\sigma} c_{i\sigma}$; $U > 0$ is the Hubbard parameter. 

\begin{figure}[h]
\begin{center}
\includegraphics[scale=0.45]{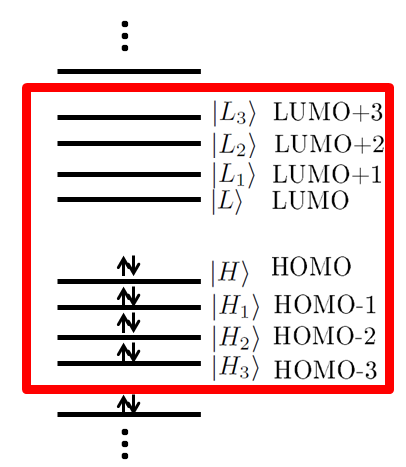}
\caption{A cartoon representing the tight binding ground state, with filled orbitals indicated by the presence of electrons (up and down arrows). The red box represents the tight binding levels involved in the diagonalization of the Hamiltonian (\ref{fulham}).}
\label{dbtAS}
\end{center}
\end{figure}

We rewrite the total Hamiltonian (\ref{fulham}) in the electron-hole basis \cite{ZSpen}; electron creation is designated by the operator $a^{\dg}_{i\sigma}$ and hole creation by $b^{\dg}_{i\sigma}$. As outlined previously, we select an active space defined by a set of tight-binding excited states. Along with the tight-binding ground state, we use those states to diagonalize the full Hamiltonian (\ref{fulham}); the electron and hole operators that are involved in writing these tight-binding excited states are indicated in Fig. \ref{dbtAS}. The ground state, within these approximations, is written as
\beq
|g\ra = c_{0}|0\ra + \sum_{mm'nn'} c_{mm';nn'} |\overline{L_{m}L_{m'};H_{n}H_{n'}}\ra,
\label{gs}
\eeq
the coefficients $c_{i}$ are complex numbers, $|0\ra$ refers to the closed shell noninteracting singlet ground state, and the states $\overline{|L_{m},L_{m'};H_{n}H_{n'}}\ra$ are double excitations \cite{ZSpen}. Here $L_{m}$ and $H_{n}$ indicate respectively an electron in the LUMO+m level and a hole in the HOMO-n level. Our approach includes states that are diradical and polyradical in character.

We choose the Hubbard parameter, $U$, such that the first singlet transition energy matches that of the literature experimental value. Since our calculations are done for isolated molecules, the most appropriate comparison would be to gas phase data. However, we are unaware of any gas phase data on the electronic states of DBT, and as a result we compare with quantum chemical calculations \cite{depr10}; the Hubbard parameter was set to $U=6.89$ eV; this value is comparable with the Hubbard parameters for the -acenes \cite{ZSpen}. 

\section{\label{eos} Electronic States and Oscillator Strengths}
We predict several low lying singlet and triplet excited states. We label all singlet transitions which are bright as $S_{n}$ where $n=1,2,3,4,5$; the transition energy for the singlet excited states, as well as their oscillator strengths of their transitions from the ground state are presented in Table \ref{S1tables}. In Table \ref{stables} we also present singlet states which are not bright. The oscillator strength is
\begin{equation}
f_{qq'} = \frac{2m_{e}\omega_{qq'}}{3\hbar e^{2}} \sum_{\beta}|\mu^{\beta}_{qq'}|^{2},
\end{equation}
with $\omega_{qq'}$ the frequency difference between states $q$ and $q'$, $\mu^{\beta}_{qq'}$ the $\beta$ component of the transition dipole matrix element between states $q$ and $q'$ and $m_{e}$ is the electron mass. The transition energy for several triplet states are presented in Table \ref{T1tables}. 


\begin{center}
\begin{table}[h]
\begin{tabular}{|c|c|c|c|}
\hline 
\textbf{State} & \textbf{Energy (eV)} & $f$ & \textbf{Direction}\tabularnewline
\hline 
\hline 
$S_{1}$ & 1.58 (1.58) & 0.955 (0.383) & $\hat{\mathbf{y}}$\tabularnewline
\hline 
$S_{2}$ & 3.31 & 0.564 & $\hat{\mathbf{y}}$\tabularnewline
\hline 
$S_{3}$ & 4.34 & 0.499 & $\hat{\mathbf{x}}$\tabularnewline
\hline 
$S_{4}$ & 4.47 & 0.197 & $\hat{\mathbf{y}}$\tabularnewline
\hline 
$S_{5}$ & 4.58 & 0.133 & $\hat{\mathbf{y}}$\tabularnewline
\hline 
$S_{6}$ & 5.31 & 0.367 & $\hat{\mathbf{y}}$ \tabularnewline
\hline
\end{tabular}
\caption{Table of singlet states with associated oscillator strengths denoted by $f$ and directions of the transition dipole moment; literature calculated values, from Deperasinska \textit{et al.} \cite{depr10}, are indicated by brackets where relevant. The molecular axes are shown in Fig. \ref{dbtmolfig}}
\label{S1tables}
\end{table}
\end{center}

\begin{center}
\begin{table}[h]
\begin{tabular}{|c|c|}
\hline 
\textbf{State} & \textbf{Energy (eV)}\tabularnewline
\hline 
\hline 
$S_{D_{1}}$ & 3.07 (2.70) \tabularnewline
\hline 
$S_{D_{2}}$ & 3.08 (2.87)\tabularnewline
\hline 
$S_{D_{3}}$ & 5.03 \tabularnewline
\hline 
\end{tabular}
\caption{Dark singlet states computed from our model; literature calculated values, from Deperasinska \textit{et al.} \cite{depr10}, are indicated by brackets where relevant}
\label{stables}
\end{table}
\end{center}

\begin{center}
\begin{table}[h]
\begin{tabular}{|c|c|}
\hline 
\textbf{State} & \textbf{Energy (eV)}\tabularnewline
\hline 
\hline 
$T_{1}$ & 0.83 (0.23)\tabularnewline
\hline 
$T_{2}$ & 2.24 (1.96)\tabularnewline
\hline 
$T_{3}$ & 2.32 (2.13)\tabularnewline
\hline 
$T_{4}$ & 2.57 (2.40)\tabularnewline
\hline 
$T_{5}$ & 2.71 (2.65)\tabularnewline
\hline 
\end{tabular}
\caption{Table of triplet states; literature calculated values of triplet energies, taken from Deperasinska \textit{et al.} \cite{depr10}, are indicated by brackets}
\label{T1tables}
\end{table}
\end{center}

\begin{center}
\begin{table}[h]
\begin{tabular}{|c|c|}
\hline 
\textbf{Natural Orbital} & \textbf{Occupation Number}\tabularnewline
\hline 
\hline 
HONO-1 & 1.974\tabularnewline
\hline 
HONO & 1.904\tabularnewline
\hline 
LUNO & 0.0972\tabularnewline
\hline 
LUNO+1 & 0.0365\tabularnewline
\hline 
\end{tabular}
\caption{Natural orbital occupation numbers (NONOs) for various natural orbtials in DBT.}
\label{HONO}
\end{table}
\end{center}

We find reasonable quantitative agreement with energies from more sophisticated quantum chemical calculations \cite{depr10} for the dark singlet states as well as higher energy triplet states; however, our model significantly disagrees with the value of  Deperasinska \textit{et al.} for the first triplet state. They predict this state has an energy of 0.23 eV above the ground state; we predict 0.83 eV, comparable to that of the mid sized -acenes. We find the first doubly excited state, $2LH$, has an energy of 1.99 eV above the ground state; this $2LH$ state is composed mainly of two electrons being excited from the HOMO state to the LUMO state. 

We have also computed natural orbital occupation numbers for DBT. The occupation of the highest occupied natural orbital (HONO), the lowest unoccupied natural orbital (LUNO), as well as HONO-1, and LUNO-1 are computed and presented in Table \ref{HONO}. We can see that HONO, HONO-1 (LUNO, LUNO+1) occupation is very close to 2 (0), indicating that the ground state can be considered to be mostly closed shell singlet in character; this is in contrast to the -acenes, where in the large acene limit the ground state is essentially polyradical in character.

\section{\label{sos} The $S_{1}$ and $T_{1}$ States}
\subsection{\label{HP} Energy scaling with $U$}

%

Upon diagonalization of the total Hamiltonian (\ref{fulham}) within our active space \cite{ZSpen}, the singlet and triplet states can be expressed as a superposition of tight-binding single excitations; single excitations are composed of one electron and one hole creation operator acting on the tight-binding ground state. The first singlet and triplet excited states can be written as 
\beq
\label{s1eq}
&& |S_{1}\ra = \sum_{mn\sigma} c_{mn\sigma;S_{1}} |\overline{L_{m},H_{n};\sigma}\ra,\\
&& |T_{1}\ra = \sum_{mn\sigma} c_{mn\sigma;T_{1}} |\overline{L_{m},H_{n};\sigma}\ra, \label{t1eq} 
\eeq
where $|\overline{L_{m},H_{n};\sigma}\ra$ indicates a tight-binding state with an electron of spin $\sigma$ removed from the HOMO-n state and placed in the LUMO+m state; $c_{nm\sigma;X}$ is the amplitude of the tight-binding excitation $|\overline{L_{m},H_{n};\sigma}\ra$ for the state $X$. The expectation values of the energies of these states are


\beq
\label{s1HP}
&& E_{S_{1}} = \la S_{1} | H_{Hu} | S_{1} \ra = E_{0;S_{1}} + U\sum_{\alpha} \tilde{\Gamma}_{S_{1},\alpha}, \\
&& E_{T_{1}} = \la T_{1} | H_{Hu} | T_{1} \ra = E_{0;T_{1}} + U\sum_{\alpha} \tilde{\Gamma}_{T_{1},\alpha}, 
\label{t1HP}
\eeq
where the index $\alpha$ runs over the sites. The term $E_{0;X}$ is
\beq
E_{0;X} = E_{0} + \sum_{nm\sigma} |c_{mn\sigma;X}|^{2} \hbar\omega_{L_{m}H_{n}}.
\eeq
The term $E_{0}$ is the sum over the energies of all the filled levels in the tight-binding ground state; $\hbar\omega_{L_{m}H_{n}}$ is the difference in energy between two tight-binding eigenstates $L_{m}$ and $H_{n}$. Thus the term $E_{0;X}$ has an implicit dependence on $U$ from the coefficients $c_{mn\sigma;X}$; however, this term does not vary much for $X=S_{1},T_{1}$ for $U$ varying from zero to the values adopted in this work. For a particular state $X$, the term $\tilde{\Gamma}_{X,\alpha}$ is 
\beq
\tilde{\Gamma}_{X,\alpha} = \sum_{\substack{nm\sigma, \\ n'm'}}c_{nm\sigma;X} c^{*}_{n'm'\tilde{\sigma};X} \tilde{\Gamma}_{L_{n'}H_{m'}H_{m}L_{n};\alpha},
\label{HPdef}
\eeq
where
\beq
\tilde{\Gamma}_{L_{n'}H_{m'}H_{m}L_{n};\alpha} = M_{L_{n'},\alpha}M^{*}_{H_{m'},\alpha}M_{H_{m},\alpha}M^{*}_{L_{n},\alpha}.
\label{gamdef}
\eeq
Here $M_{k,\alpha}$ is the amplitude of the $\alpha^{th}$ site for a tight-binding state $k$;  these quantities were defined in our earlier work \cite{ZSpen}.  The term $\tilde{\Gamma}_{L_{n'}H_{m'}H_{m}L_{n};\alpha}$ describes the overlap of the electron and hole wavefunctions for two states $|\overline{L_{n},H_{m};\sigma}\ra$ and $|\overline{L_{n'},H_{m'};\tilde{\sigma}}\ra$ for a particular site $\alpha$. The sum over all sites of $\tilde{\Gamma}_{X,\alpha}$ multiplied by the Coulomb parameter, $U$, gives the adjustment of the energy of the state; again there is some implicit $U$ dependence in the coefficients $c_{nm\sigma;X}$, but it is not significant.

\begin{center}
\begin{table}[h]
\begin{tabular}{|c|c|c|c|c|}
\hline 
$\sum_{\alpha}\tilde{\Gamma}_{X,\alpha}$ & \textbf{DBT} & \textbf{Tetracene} & \textbf{Pentacene} & \textbf{Hexacene}\tabularnewline
\hline 
\hline 
$S_{1}$ & 0.0449 & 0.0734 & 0.0626 & 0.0519\tabularnewline
\hline 
$T_{1}$ & -0.0659 & -0.137 & -0.122 & -0.113\tabularnewline
\hline 
HOMO-LUMO Ref. & 0.0537 & 0.105 & 0.0927 & 0.0835\tabularnewline
\hline 
\end{tabular}
\label{tilgam1}
\caption{The quantity $\sum_{\alpha} \tilde{\Gamma}_{X,\alpha}$ computed for $X = S_{1},T_{1}$ and a reference for an excitation from the HOMO level to the LUMO level, for DBT and the -acenes.}
\end{table}
\end{center}

The energy adjustment from the inclusion of the Hubbard term depends mainly on the degree of overlap of the tight-binding electron and hole wavefunctions. For $\sum_{\alpha}\tilde{\Gamma}_{X,\alpha} > 0$ the energy of the state increases with increasing Coulomb repulsion $U$, as is the case for singlet states, and for $\sum_{\alpha}\tilde{\Gamma}_{X,\alpha} < 0$ the energy of the state decreases with increasing Coulomb repulsion, as is the case for triplet states; the magnitude of $\sum_{\alpha}\tilde{\Gamma}_{X,\alpha}$ indicates the amount by which a state is affected by the introduction of Coulomb repulsion. The quantity $\sum_{\alpha} \tilde{\Gamma}_{X,\alpha}$ for DBT and the -acenes is presented in Table V. While the ratio of the correction of the triplet state to the singlet state remains the same in both DBT and the -acenes, the magnitude of the correction factors for DBT, $\sum_{\alpha} \tilde{\Gamma}_{S_{1},\alpha}, \sum_{\alpha}\tilde{\Gamma}_{T_{1},\alpha}$ is  different from those of tetracene, a molecule of similar structure, by approximately a factor of two. This leads to a smaller singlet-triplet gap in DBT than in the -acenes. 

The low singlet-triplet gap in DBT can be attributed to the relatively high delocalization of the tight-binding eigenstates, compared with the mid sized -acenes. This delocalization is reflected in the term $\tilde{\Gamma}_{X,\alpha}$ (\ref{HPdef}); the expectation values for energy of the singlet and triplet excited states (\ref{s1HP},\ref{t1HP}) are proportional the sum of (\ref{HPdef}) over all sites, which is proportional to the overlap of the electron and hole wavefunctions of the tight-binding excitations that make up these states (\ref{gamdef}). These overlaps are small for DBT.


\subsection{\label{dcorr} Charge Density Correlation Function}
To characterize the impact of the Hubbard interaction on the states of interest we compute the density correlation function, denoted by $g^{(2)}(s)$, for these states. This correlation function is covered in detail in our previous work \cite{ZSpen}; it indicates the correlation between the electrons in these states, and is a discrete version of the electron density correlation function used in condensed matter physics. We construct $g^{(2)}(s)$, by averaging the site density correlation function over all pairs $(i,j)$ of sites in the molecule with the same bond length distance between them \cite{ZSpen}. 

\begin{figure}[h]
\begin{center}
\includegraphics[scale=0.45]{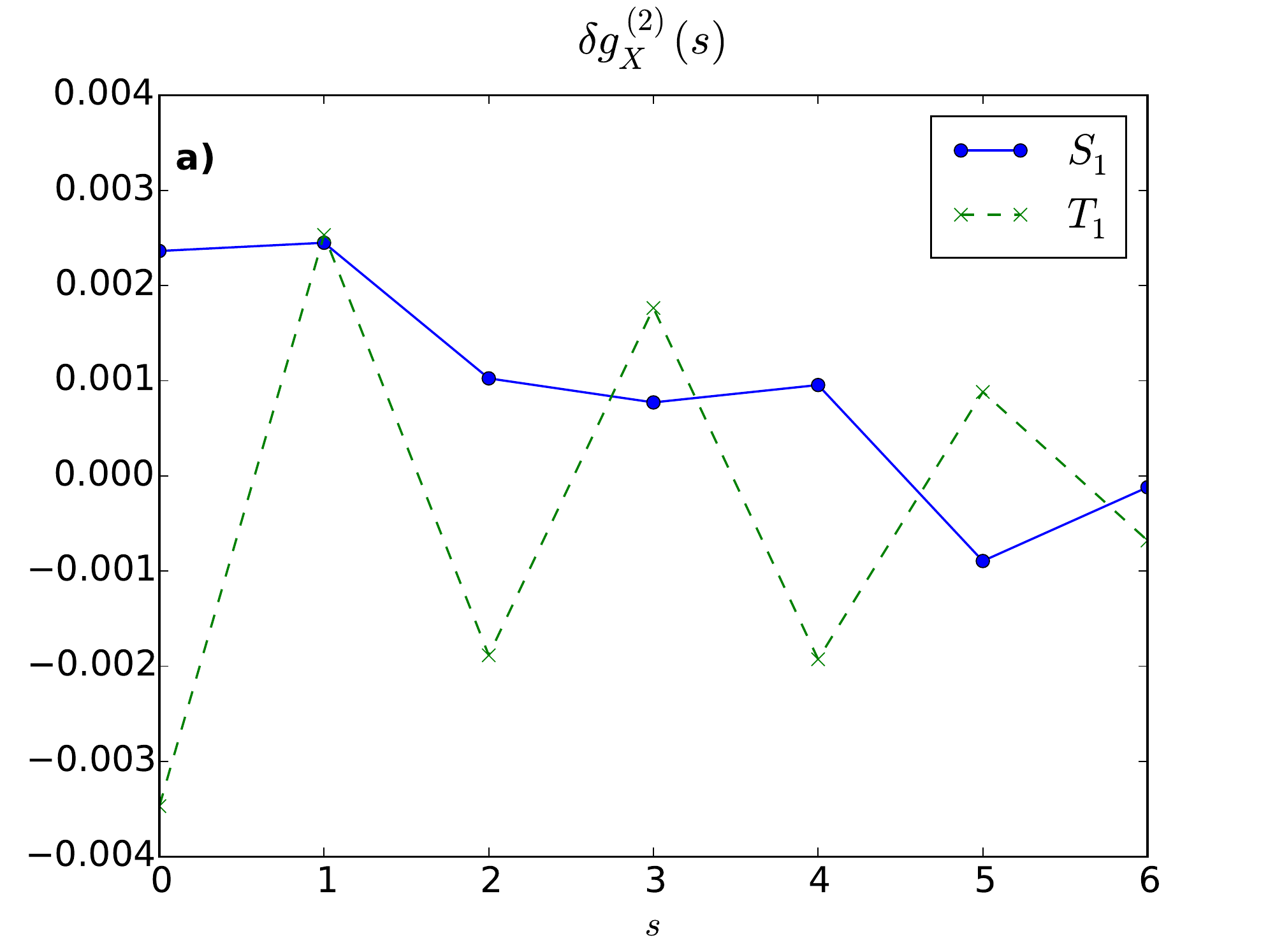}
\includegraphics[scale=0.45]{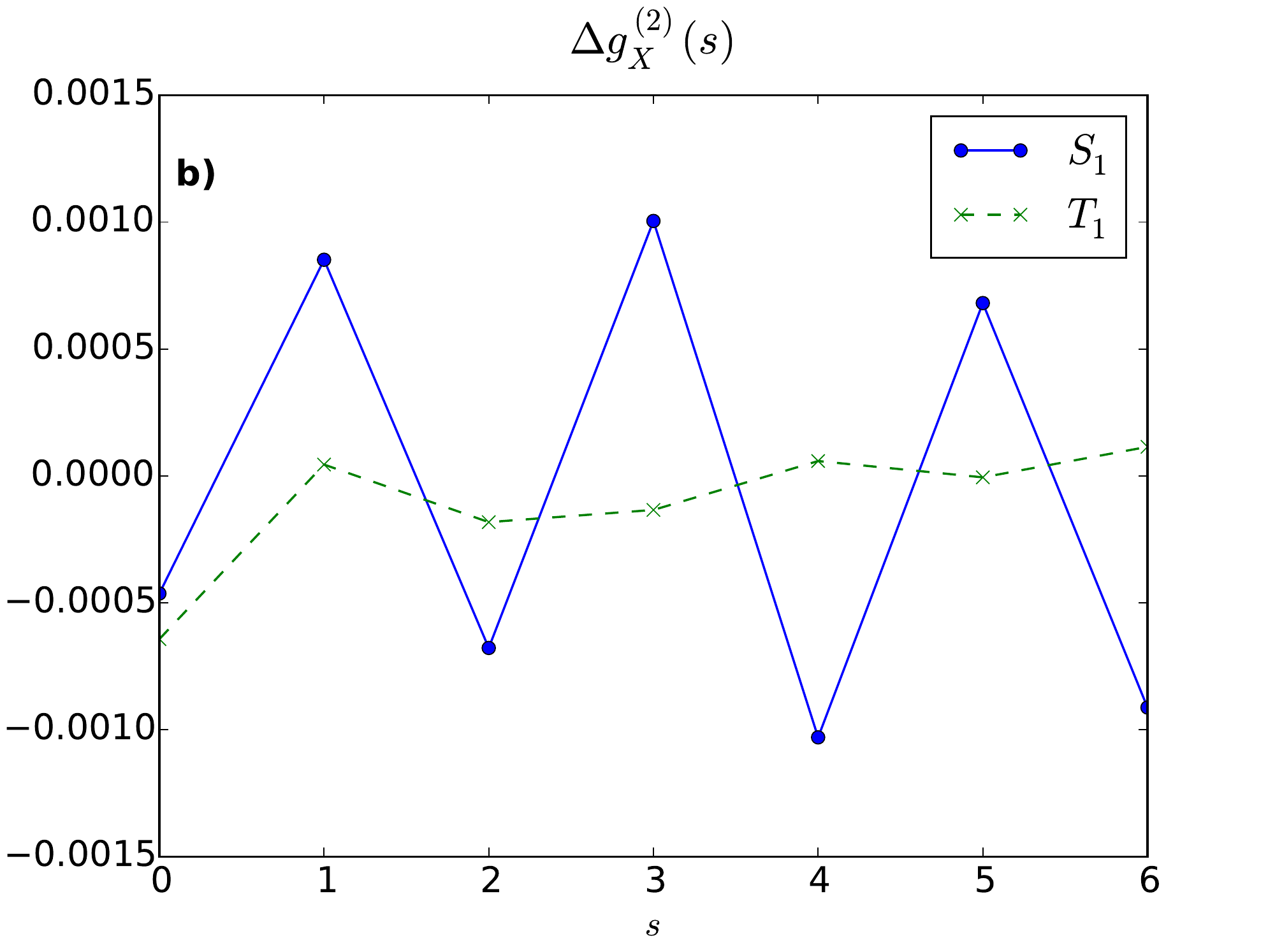}
\caption{a) The $\delta g^{(2)}_{X}$ function is plotted for $X = S_{1}$ and $T_{1}$. The gap between $\delta g^{(2)}_{X}(0)$ is around four times smaller than the equivalent gap in the -acenes \cite{ZSpen}, resulting in a smaller singlet-triplet splitting than in the -acenes. b) The $\Delta g^{(2)}_{X}$ function is plotted for $X = S_{1}$ and $T_{1}$; upon the introduction of the Coulomb repulsion (\ref{HubHam}), the Fermi hole deepens for both the singlet and triplet states. The $\Delta g^{(2)}_{X}(0)$ is also around four times smaller than the equivalent quantity in the -acenes; this relatively small adjustment in the electron density correlation is related to the small singlet-triplet gap in DBT.}
\label{s1t1g1}
\end{center}
\end{figure}

We calculate this quantity for states of interest where the Hubbard Hamiltonian (\ref{HubHam}) is included, which we denote by $g^{(2)}_{Hu;X}$, and where the Hubbard Hamiltonian is neglected which we denote as $g^{(2)}_{TB;X}(s)$. In this section we focus on the first two singly excited states: the first singlet state $S_{1}$, and the first triplet $T_{1}$. To discern the difference between electron correlation in these states in their tight-binding limit from that in the tight-binding ground state we define
\beq
\delta g^{(2)}_{X} = g^{(2)}_{TB;X} - g^{(2)}_{TB;GS},
\label{dg1}
\eeq
where $X$ identifies the tight-binding equivalent of $S_{1}$ and $T_{1}$, and $g^{(2)}_{TB;GS}(s)$ refers to computing $g^{(2)}(s)$ for the tight-binding ground state; this is plotted in Fig. \ref{s1t1g1} a). To quantify the impact of the Hubbard Hamiltonian on these states, we define
\beq
\Delta g^{(2)}_{X}(s) = g^{(2)}_{Hu;X}(s) - g^{(2)}_{TB;X}(s),
\label{dg2}
\eeq
where $g^{(2)}_{Hu;X}(s)$ indicates $g^{(2)}(s)$ computed for a particular state $X = S_{1}, T_{1}$, and $g^{(2)}_{TB;X}(s)$ refers to $g^{(2)}(s)$ computed for the equivalent tight-binding state; this is plotted in Fig. \ref{s1t1g1} b). 

Typically, singlet (triplet) states have a spatial component of their wavefunction symmetric (antisymmetric) with respect to exchange of particle coordinates, leading to a larger (smaller) spatial overlap of electrons in the singlet (triplet) states; in Fig. \ref{s1t1g1} a) it is clear that $\delta g^{(2)}_{S_{1}}(0) > 0$ ($\delta g^{(2)}_{T_{1}}(0) < 0$), indicating a shallower (deeper) Fermi hole relative to the tight-binding ground state. The difference between $\delta g^{(2)}_{S_{1}}(0) - \delta g^{(2)}_{T_{1}}(s)$ is around four times smaller than that of the -acenes, indicating that the difference in electron overlap in the DBT $S_{1}$ and $T_{1}$ states is far less than the corresponding electron overlap in the -acene $S_{1}$ and $T_{1}$ states, hence the relatively small singlet-triplet gap in DBT. 

Upon the introduction of Coulomb repulsion (\ref{HubHam}), the energy degeneracy of the singlet and triplet states is lifted; this interaction also modifies the electron motion, and we find that the Fermi holes are deeper for the interacting $S_{1}$ and $T_{1}$ states relative to their noninteracting equivalents. The value of $\Delta g^{(2)}_{S_{1}}(0)-\Delta g^{(2)}_{T_{1}}(0)$ for DBT is again around four times smaller than the equivalent value in the -acenes; such a relatively small correction to the density correlation is related to the relatively small singlet-triplet gap.

This analysis is also carried out for the first doubly excited state, $2LH$, in Appendix \ref{dos}.

\section{\label{conc} Conclusion}


We have used a computationally and physically simple scheme, involving a Hubbard model with a limited basis, to extract the electronic excited state energies, oscillator strengths, and wavefunctions of the dibenzoterrylene (DBT) molecule. We computed the HONO-LUNO occupation number for the ground state and showed that the ground state of DBT can be thought of as a mostly closed shell singlet state. We have shown that the transition energies and oscillator strengths agree reasonably well with state of the art quantum chemistry calculations; interestingly, our simple calculation suggests the existence of several bright excited states that are not predicted by a much more sophisticated DFT calculation. Our calculation for the singlet-triplet spacing is also very different from literature predictions; we predict a splitting of approximately 0.75 eV, while an earlier calculation found this to be greater than 1 eV. This relatively small gap can be attributed to the impact of the Hubbard Hamiltonian, quantified by the term $\sum_{\alpha} \tilde{\Gamma}_{\alpha;X}$ for $X=S_{1},T_{1}$; $\sum_{\alpha} \tilde{\Gamma}_{\alpha;X}$ is rather small for DBT compared with the -acenes, a set of $\pi$ conjugated molecules of a comparable structure to DBT. This adjustment, $\sum_{\alpha} \tilde{\Gamma}_{\alpha;X}$, is small in DBT due to the relatively high delocalization of the tight-binding eigenstates, compared to the -acenes. 

We then computed the density correlation function for the first singlet and triplet state in an attempt to characterize the electron behavior in these states. The singlet-triplet splitting in DBT is remarkably small, and this is exhibited in the $\delta g_{X}^{(2)}(s)$ function where the difference in electron overlap between the singlet and triplet state, exemplified by the difference in the depth of the Fermi hole in these states, is four times smaller than that of the -acenes.

\begin{widetext}
\begin{appendix}
%
\section{\label{dos} Ground and Doubly Excited State Correlation Functions}

\begin{figure}[h]
\begin{center}
\includegraphics[scale=0.45]{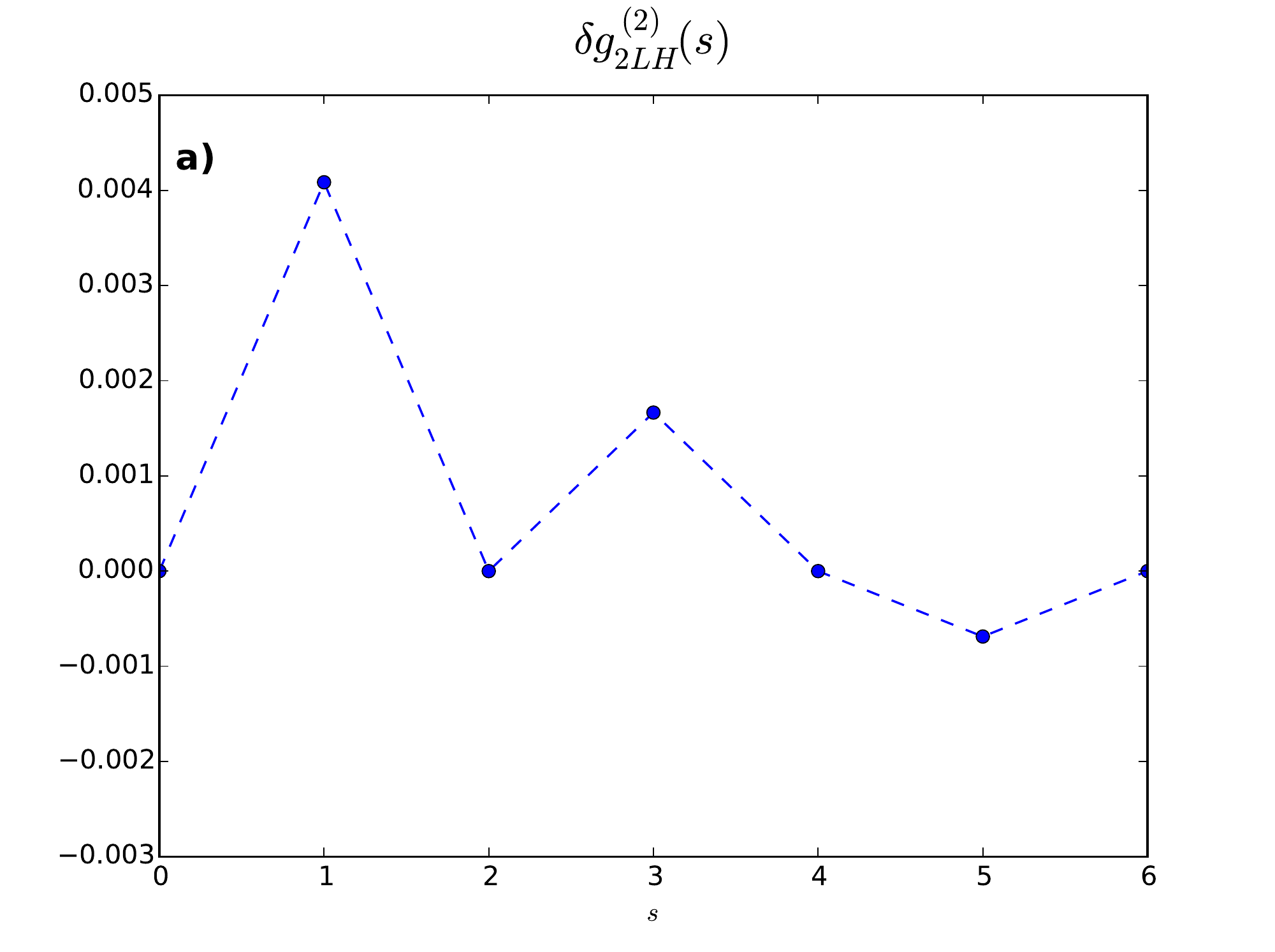}
\includegraphics[scale=0.45]{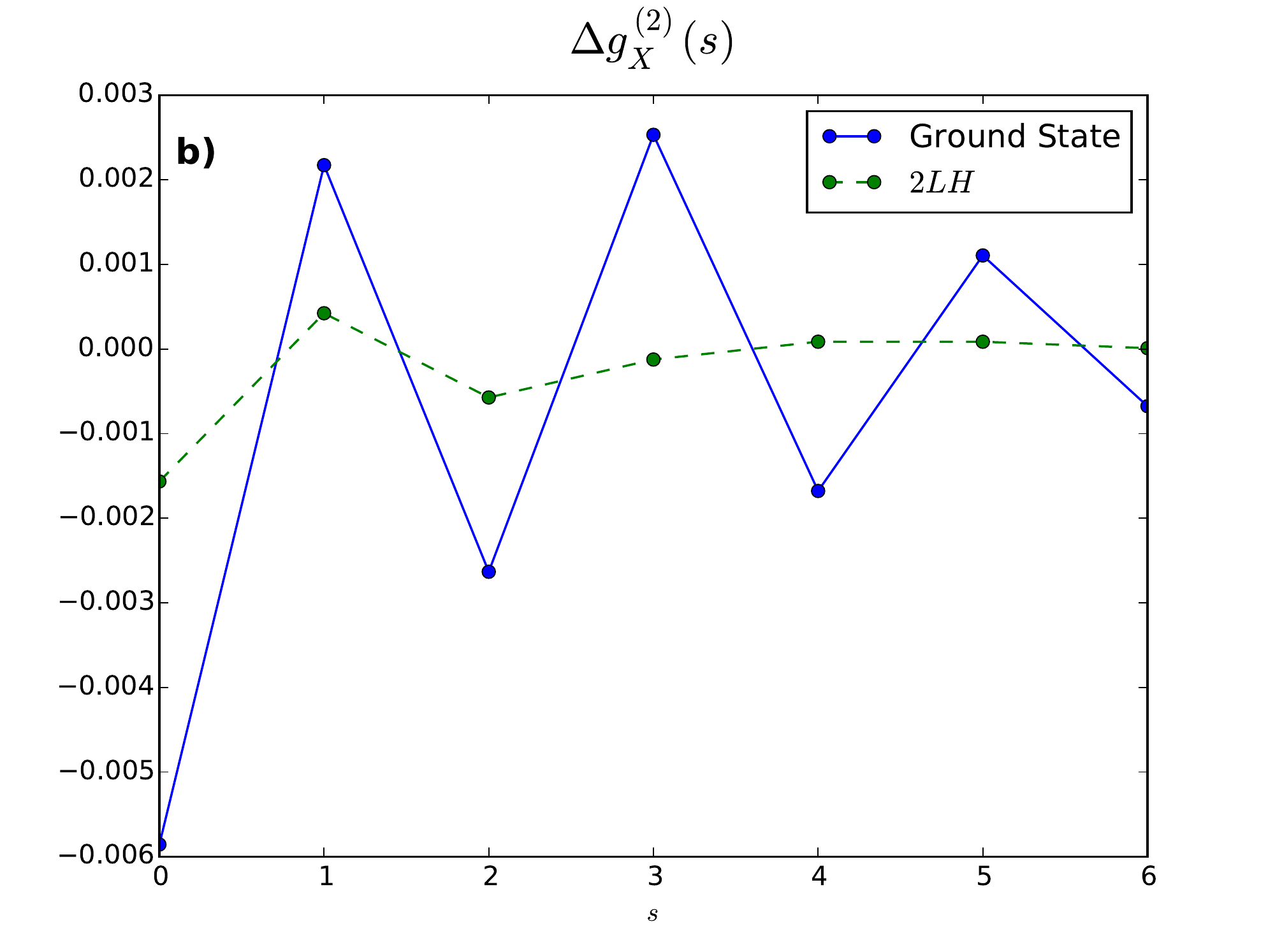}
\caption{a) The function $\delta g^{(2)}_{2LH}(s)$ is plotted and b) the function $\Delta g^{(2)}_{X}(s)$ is plotted for $X =$ ground state and $2LH$. The $\Delta g^{(2)}(0)$ for the ground and the $2LH$ state in DBT is around four times smaller than that of the -acenes.}
\label{g2LH}
\end{center}
\end{figure}

We plot $\delta g^{(2)}_{2LH}(s)$ and $\Delta g^{(2)}_{X}(s)$ for the ground and the $2LH$ state in Fig. \ref{g2LH}. The $\delta g^{(2)}_{2LH}(0)$ is exactly zero, this is due to the symmetry of the tight-binding eigenfunctions \cite{ZSpen}. Upon the introduction of the Coulomb repulsion (\ref{HubHam}), the Fermi hole of both the ground state and the $2LH$ is deepened with respective to their tight-binding counterparts; this deepening of the Fermi hole is about four times smaller than the deepening of the Fermi hole in the equivalent states in the -acenes. This relatively small correction can be attributed to the a high level of delocalization in the DBT tight-binding eigenstates.
\end{appendix}
\end{widetext}
\end{document}